\documentstyle{mn}
\input{epsf}
\begin{document}
\def\p1937{PSR B1937$+$21}

\title[Giant Radio Pulses from \p1937 at 327 MHz]
{Giant Radio Pulses from the Millisecond Pulsar \p1937 at 327 MHz}

\author[M. Vivekanand]{M. Vivekanand \thanks{vivek@ncra.tifr.res.in} \\
National Center for Radio Astrophysics, TIFR,  Pune University Campus, P. O. Box 3, 
Ganeshkhind, Pune 411007, India.}

\date{}

\maketitle
\begin{abstract}
Seven giant radio pulses were recorded from the millisecond pulsar \p1937 during 
$\approx$ 8.1 minutes observation by the Ooty Radio Telescope (ORT) at 326.5 
MHz. Although sparse, these observations support most of the giant pulse behavior 
reported at higher radio frequencies (430 to 2380 MHz). Within the main component
of the integrated profile, they are emitted only in a narrow ($\la$ 47 $\mu$s) 
window of pulse phase, close to its peak. This has important implications for 
doing super-high precision timing of \p1937 at low radio frequencies.
\end{abstract}
\begin{keywords}
pulsars: general -- pulsars: individual (\p1937): stars -- neutron
\end{keywords}
\section{Introduction}

Recently Kinkhabwala and Thorsett \cite{KT2000} (henceforth KT) studied giant 
radio pulses from the fastest known millisecond pulsar to date, \p1937, 
simultaneously at frequencies 2380 mega hertz (MHz), 1420 MHz and 430 MHz, 
using total
intensity data from the Arecibo Radio Telescope (ART) (see also Sallmen \& 
Backer \cite{SB1995}; Cognard et al. \cite{C1996}). They used coherent 
dedispersion to obtain a time resolution of $\approx$ 0.1 to 0.2 micro seconds 
($\mu$s). Their main conclusions are (1) `` at each frequency, giant 
pulses are emitted only in narrow ($\la$ 10 $\mu$s) windows of pulse phase 
located $\approx$ 55-70 $\mu$s after the main and interpulse peaks'', (2) 
their `` ...  mean arrival phase appears stable'', (3) their cumulative 
intensity distribution is roughly power law with an exponent of $-1.8$.

This article reports the result of analyzing seven giant pulses from 
\p1937, obtained at 326.5 MHz using ORT; only a single linear polarization 
was available, and incoherent dedispersion was used. Although the data were
observed with a sampling interval of 102.4 $\mu$s, the pulse phases of giant 
pulses were obtained with accuracies of 15 to 20 $\mu$s; the instrument, the 
method of observation, and the several variations of data reduction for 
diverse purposes are given in Vivekanand et al. \cite{VAM1998}, Vivekanand 
\cite{MV2000}, and Vivekanand \cite{MV2001}. While the above pulse phase 
accuracies are $\la$ two orders of magnitude lower than those of KT, 
they are sufficient for the current purpose; they are much smaller than
the widths of giant pulses that are affected by interstellar scattering 
and dispersion broadening at 326.5 MHz (discussed later).

Table 1 lists 5 of the total 7 data files observed during the later half of 
May 1995, in which giant pulses were found. For technical reasons, the first 
6\,000 of the 681\,984 samples (each of duration 102.4 $\mu$s) were ignored 
in each file. The data were folded at the corresponding period (table 1); 15 
times samples were synthesized in each period, whose duration differed from 
file to file, but is approximately 103.85 $\mu$s. In each file, consists of 
44\,436 useful periods of data, the five highest peaks were found, and giant 
pulses were identified as those whose peak signal to noise ratio (SNR) was 
higher than or close to 10.0. Then their pulse phases and their uncertainties 
were derived as in Vivekanand \cite{MV2001}. Then the integrated profile of
each file was cross correlated with that of file observed at UT 22:24:37 on 
1995-05-22 (no giant pulses were found in this file), and the data were 
shifted in each file (using FFT technique) by the corresponding amount of 
time.

\begin{table}
\begin{tabular}{@{}c@{\ \ }c@{\ \ }c@{\ \ }r@{\ \ }r@{\ \ }c}
\hline
\hline
DATE & UT & PERIOD & SNR & POS & ERR \\
\hline
\hline
1995-05-16 & 22:11:28 & 1.55770189 &  9.11 & 0.244 & 0.010 \\
1995-05-16 & 22:13:45 & 1.55770191 &  8.43 & 0.235 & 0.015 \\
1995-05-22 & 22:26:31 & 1.55770736 & 18.44 & 0.774 & 0.009 \\
           &          &            & 17.52 & 0.231 & 0.010 \\
           &          &            & 10.88 & 0.248 & 0.014 \\
1995-05-22 & 22:28:17 & 1.55770738 & 14.20 & 0.244 & 0.010 \\
1995-05-22 & 22:30:06 & 1.55770740 & 14.85 & 0.248 & 0.010 \\
\hline
\end{tabular}
\caption{
	Columns 1 and 2 contain the date and time (UT) of observation; column 
	3 contains the period (in ms) used for folding, derived from the TEMPO 
	package; column 4 contains the peak signal-to-noise ratio (SNR) of the 
	giant pulse, while the last two columns contain the pulse phases of 
	giant pulses and their errors (in units of phase within the period).
        }
\end{table}

\begin{figure}
\epsfxsize=8.5cm \epsfbox{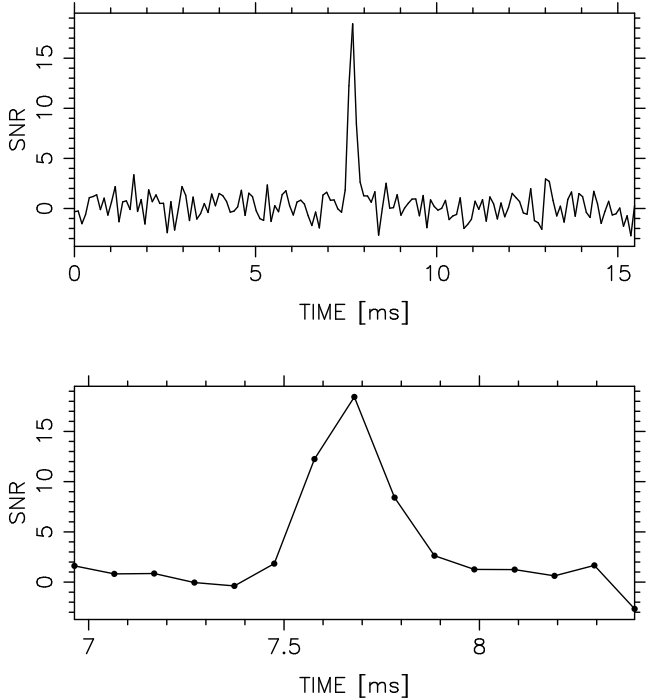}
\caption{
         {\bf Top panel}: Ten periods of data centered on the strongest spike 
	 observed in \p1937 at 326.5 MHz using ORT (file observed at UT 22:26:31 
	 on 1995-05-22). The abscissa is in milli seconds, while the ordinate is 
	 in units of the signal to noise ratio. {\bf Bottom panel}: The period 
	 containing the giant pulse.
        }
\label{fig1}
\end{figure}

\p1937 has a very small rotation period; so it requires fast sampling rates that are 
not available at ORT. It has a high dispersion measure; so coherent dedispersion 
technique is required to maintain the required sensitivity; that is also not available 
at ORT. Therefore observing this pulsar at ORT is a difficult exercise, although ORT 
is the most powerful telescope operating currently at $\approx$ 327 MHz. The top panel 
of fig.~\ref{fig1} shows ten periods of \p1937 centered around the strongest giant pulse 
observed at ORT. The ordinate was obtained in units of the signal to noise ratio as 
described in Vivekanand \cite{MV2001}. The effects of dispersion smearing and interstellar 
scattering are evident in the bottom panel of fig.~\ref{fig1}.

\section{Intensity Distribution}

These data are too meager to verify the cumulative intensity distribution reported 
by KT of giant pulses of \p1937. So only a consistency check will be done here, 
using their 430 MHz observations. This frequency is appropriate for comparison 
with current ORT observations since interstellar scattering of giant pulses affects 
their 430 MHz observations, just as it does the ORT observations, but hardly their 
1420 MHz or 2380 MHz observations. 

The effective collecting area of ART is $\approx$ 49\,700 square meters (m$^2$),
obtained from the antenna gain $G$ value of 18.0 Kelvin per Jansky (K/Jy) quoted by
Thorsett \& Stinebring (\cite{TS1990}) for observations of \p1937. That of ORT is
$\approx$ 7\,400 (m$^2$); this is obtained from the $G = 1.9$ (K/Jy) (Selvanayagam et 
al \cite{SPN1993}) for the correlation mode of operation of ORT, and noting that the 
current observations have been done in the total power mode, which gains a factor 
$\sqrt{2}$. This reduces by factor $\approx$ 6.7 the sensitivity of ORT with respect 
to that of ART. The system temperature of ART is 170 K towards \p1937 (Thorsett \& 
Stinebring \cite{TS1990}); that of ORT is the receiver temperature plus spillover of 
110 K (Selvanayagam et al \cite{SPN1993}) plus the galactic background contribution 
of 202 K, obtained by scaling 117 K (see equation 8-6 of Manchester \& Taylor 
\cite{MT1977}) by the frequency scaling factor $(400/326.5)^{2.7} \approx 1.73$ (see 
Salter \& Brown \cite{SB1988} for the exponent 2.7). This reduces ORT's relative 
sensitivity by the further factor $312/170 \approx 1.84$. The total available 
bandwidths at the two telescopes contribute a factor $\sqrt{10/8} \approx$ 1.12, 
while a factor $\sqrt{2}$ arises due to the availability of only a single 
polarization at ORT. 

At 430 MHz, KT look for giant pulses by integrating the flux density within a 150 
$\mu$s window ``located on the tails of the main pulse (MP) and the interpulse (IP)''; 
in this work they are located by means of their peak SNR, measured within a single 
time sample of duration 102.4 $\mu$s. Further, interstellar scattering broadening of 
giant pulses in KT is $\approx$ 30 $\mu$s at 430 MHz (their table 1), while 
interstellar dispersion smearing is negligible due to coherent dedispersion. Thus 
at ART the energy of giant pulses can be assumed to be effectively spread over the 
detection window width of $ 150 \mu$s. On the other hand, scatter broadening at ORT 
is $\approx 30 \times \left ( 430.0 / 326.5 \right )^4 \approx 90.0$ $\mu$s (see 
Manchester \& Taylor \cite{MT1977}); and dispersion smearing within each frequency 
channel is $\approx$ 165 $\mu$s. Now, scatter broadening occurs at each frequency 
within the observing bandwidth. Thus at ORT the energy of giant pulses can be 
expected to be spread over $\approx \sqrt{90^2 + 165^2} \approx$ 188 $\mu$s. Thus 
the relative sensitivity of ORT is reduced by $\approx 188 / 150 \approx 1.25$ due 
to smearing of giant pulse energy over a longer time, and by $\sqrt{150.0/102.4} 
\approx 1.21$ due to the reduced measurement window.

ORT's relative sensitivity is further affected by two uncertain factors. The first 
is a reduction by factor $\alpha$, due to refractive interstellar scintillation, 
phasing of ORT, etc; $\alpha$ can be $\approx$ 2 or even more at ORT (see 
Vivekanand \cite{MV1995}).

The second is $\beta$, due to the relative SNR thresholds chosen in both works. 
The threshold is $\approx$ 10 in this work; i.e., at ORT one has observed $7$ 
giant pulses in $\approx 8.1$ minutes of observation above a SNR threshold of 
$\approx 10$. In KT this number is not explicitly stated. However, from their 
fig.~7 one can deduce that a fraction of $\approx 10^{-4}$ of their $\approx 
10^6$ total periods (KT mislabel these as ``pulses'' in the second last para of
page 365, which can be confused with the giant pulses) have integrated fluxes 
greater than $\approx 2000 \mathrm{\ [Jy-}\mu\mathrm{s]}$, at 430 MHz, within 
the $150 \mu$s window. The corresponding flux density threshold will be $\approx 
2000 \mathrm{\ [Jy-}\mu\mathrm{s] \ } /\ 150 \mathrm{\ [}\mu\mathrm{s]} \approx 
13\mathrm{\ [Jy]}$. Now, the noise of ART within the $150 \mu$s window would be 
$170 \mathrm{\ [K]\ } /\ G \mathrm{\ [K/Jy]\ } /\ 2\ /\ \sqrt{10 \mathrm{\ 
[MHz]\ } \times 150 \mathrm{\ [}\mu\mathrm{s]\ }} \approx 0.12$ Jy (the extra 
factor of 2 in the denominator arises due to the definition of the gain $G$ of 
an antenna). Thus KT would have observed $\approx 100$ giant pulses above the
SNR threshold of $\approx 110$ (Kinkhabwala \cite{AK2001}).

Thus the relative sensitivity of ORT is enhanced by a factor $\beta \approx 110 
/ 10 \approx 11$ due to the relative SNR thresholds of both works. Finally, the 
giant pulse flux density scales with observing frequency as $\nu^{-3.1}$ (KT); 
this would further enhance the sensitivity of ORT with respect to that of ART 
by the factor $\approx \left ( 430.0 / 326.5 \right )^{3.1} \approx$ 2.35. 

This leads to an effective sensitivity of ORT that is a factor $6.7 \times 1.84 
\times 1.12 \times 1.41 \times 1.25 \times 1.21 \times \alpha / 11 / 2.35 
\approx 1.14 \times \alpha$ less than that of ART. 

Now, KT claim that the fraction of giant pulses above a flux density $S$ is 
proportional to $S^{-1.8}$; actually the abscissa in their fig.~7 is intensity 
$I$, which is $S$ multiplied by a fixed duration of time. So the number of 
giant pulses ORT should have recorded in half an hour would be $\approx 100 \times 
\left ( 1.14 \times \alpha \right ) ^{-1.8}$. Therefore in the $\approx$ 8.1 
minutes of actual observation, ORT is expected to record $\approx 21 / \left ( 
\alpha \right ) ^{1.8}$ giant pulses.

For $\alpha \approx 1$, this is a factor $\approx 3$ higher than the number of giant 
pulses actually observed at ORT; $\alpha$ needs to be $\approx 1.8$ for the numbers 
to match, which is not unreasonable at ORT. Furthermore, small changes in some of the 
above numbers can cause significant change in the final result. For example, assuming 
that the scatter broadening is 40 (instead of 30) $\mu$s further reduces the number
of giant pulses potentially observable by $\approx 14$\%. As another example, if the 
giant pulse flux density is assumed to scale with observing frequency as $\nu^{-2.6}$,
instead of as $\nu^{-3.1}$, since such a possibility exits in fig.~10 of KT, the 
number of giant pulses potentially observable decreases by $\approx 28$\%. It is 
therefore concluded that the seven giant pulses of \p1937 observed in about 
8.1 minutes of the current ORT observations are consistent with the cumulative 
intensity distribution of KT.

\section{Phase Distribution}

The effective time resolution of these data is so many times worse than that 
of KT that once again a detailed comparison is not possible with their work.
However fig.~\ref{fig2} shows that there is broad agreement.

\subsection{Mean Arrival Phase}

\begin{figure}
\epsfxsize=8.25cm \epsfbox{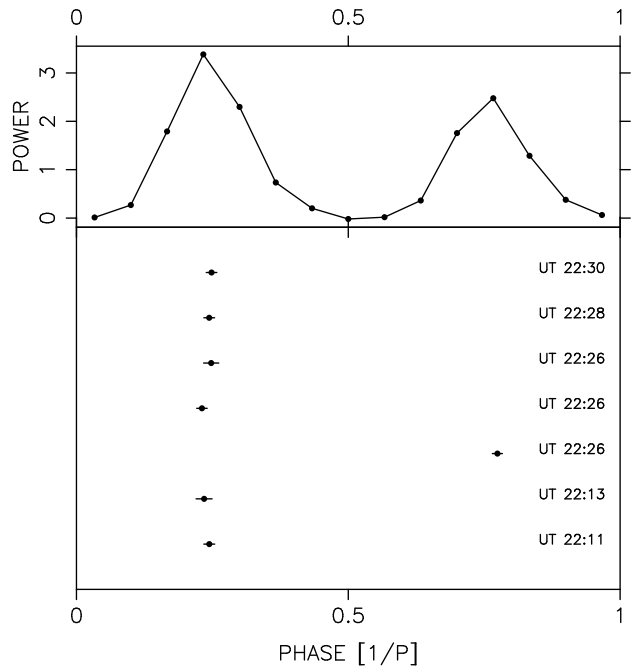}
\caption{
	Abscissa is the phase within the integrated profile, in units of
	the pulsar period. The top panel shows the integrated profile 
	(in arbitrary units) of \p1937 for the $\approx$ 311\,052 periods 
	of useful data in the seven files observed. The bottom panel shows 
	the pulse phases of the seven strongest giant pulses observed
	in five of those files (table 1). The error bars represent two 
	standard deviations. This figure should be compared with fig.~4 of 
	KT.
        }
\label{fig2}
\end{figure}

The mean pulse phase of the 6 giant pulses in the main pulse of the 
integrated profile of \p1937 (fig.~\ref{fig2}) is 0.242$\pm$0.006, in units 
of the pulsar period ($\approx$ 1.55770 ms). For reference, the phase of 
the peak of the main pulse in fig.~\ref{fig2} is 0.247 (discussed later).
The error on the mean pulse phase is $0.006 \times 1557.7 \approx 9.3 \mu$s.
This is obtained by taking into account (1) the formal estimation error on 
the phase of each giant pulse (column 6 of table 1; see Vivekanand 
\cite{MV2001} for details), (2) the departure of the estimated phase of 
each giant pulse from the mean of all six values (column 5 of table 1), 
and (3) reducing by $\sqrt{6}$ the final answer. Assuming Gaussian statistics 
one can conclude that most giant pulses (in the main pulse of \p1937 at 326.5 
MHz) arrive within a phase window of $\approx 9.3 \times 5 \approx 47 \mu$s;
this is consistent with the narrow window of $\le$ 10 $\mu$s obtained by KT
at higher radio frequencies. A corresponding comment can not be made regarding 
the giant pulses in the inter pulse of \p1937 at 326.5 MHz, since these data 
contain only one such; it occurs $\approx 19 \mu$s after the inter pulse
peak.

A visual inspection of fig.~\ref{fig2} shows that the arrival phases of 
giant pulses in the main pulse of \p1937 at 326.5 MHz are ``stable'' as 
claimed by KT. Once again this work can not comment on the stability of 
the arrival phases of giant pulses in the inter pulse.

\subsection{Its Separation From Pulse Peak}

The mean pulse phase of the peak of the main pulse of \p1937 in fig.~\ref{fig2},
obtained by weighting the abscissa with the ordinate in the top panel, is
0.247 with (comparatively) negligible error. So it is separated from the
mean pulse phase of giant pulses by $\left ( 0.242 - 0.247 \right ) \times 
1557.70 \approx - 8 \mu$s, which is within measurement errors. This implies 
that the mean arrival phase of giant pulses is coincident with the mean 
pulse phase of the main component of the integrated profile of \p1937 at 326.5 
MHz. The separation of $\approx -8 \mu$s is independent of the scatter
broadening at ORT; that is expected to bias both the mean pulse phase of a giant 
pulse (modeled as a delta function convolved with a truncated exponential 
function), as well as the mean pulse phase of the main pulse of the integrated 
profile (modeled as a Gaussian convolved with a truncated exponential 
function), by the same amount, viz., the scatter broadening width.

This is inconsistent with the $\approx 55 - 70 \mu$s separation noted by KT.
They mention that the separation has a ``lower limit of 49 $\mu$s at 430 MHz'';
this also appears to be the case from a visual inspection of the bottom panel 
of their fig.~2. Therefore the best explanation currently is that ORT 
observations, obtained using only a single linear polarization, might bias 
the mean pulse phase of the peak of the integrated profile to a greater 
extent than they do the mean pulse phases of giant pulses. This is plausible 
if, for example, the mean position angle of the linear polarization vector 
is different in the two cases; or if it changes across the integrated profile 
at a rate that is different for the rate of change across the giant pulses, 
since the former is expected to be dominated by the non-giant pulse (or the 
so called normal) emission. However this result merits verification by 
future observations.

This analysis is obviously not possible for the interpulse of \p1937 in
the current data.

\section{Discussion}

This work supports, at 326.5 MHz, the reported behavior of giant pulses of 
\p1937 at higher radio frequencies (KT), although with meager data. It is 
encouraging that this was possible with data having only a single linear 
polarization, and which also suffer from severe interstellar scattering.

KT showed that timing the giant pulses of \p1937 gives precision comparable 
to that obtained by timing the integrated profile, on the short time scales
($\approx$ hours). Whether this leads to better long term timing precision
depends upon the emission characteristics of giant pulses -- how fast does
their average profile stabilize, etc. (KT). This problem can be studied with 
actual data, as is currently being undertaken by KT at the higher radio 
frequencies. Unfortunately this can not be done at 326.5 MHz because of the 
technical limitations of ORT. It would have been nice if one could 
theoretically study the problem; but that would require more information on 
the intrinsic scatter of the giant pulse emission phase.

\section{Acknowledgements}

I thank the referee for corrections and suggestions, and Ali Kinkhabwala 
for clarifications. This research has made use of NASA's Astrophysics 
Data System Abstract Service.

\vfill
\eject
\end{document}